\documentstyle[prb,preprint,aps]{revtex}
\tightenlines
\begin{document}
\draft
\title
{\bf Temperature and magnetic field dependence of
superconductivity in nanoscopic metallic grains}
\author{\"{O}. Bozat and Z. Gedik}
\address{
Department of Physics, Bilkent University, Bilkent 06533, Ankara,
Turkey.}
\begin{titlepage}
\maketitle
\begin{abstract}
We study pairing correlations in ultrasmall superconductor in the
nanoscopic limit by means of a toy model where electrons are
confined in a single, multiply degenerate energy level. We solve
the model exactly to investigate the temperature and magnetic
field dependence of number parity effect (dependence of ground
state energy on evenness or oddness of the number of electrons).
We find a different parity effect parameter to critical
temperature ratio ($\simeq$4 rather than 3.5) which turns out to
be consistent with exact solution of the BCS gap equation for our
model. This suggest the equivalence between the parity effect
parameter and the superconducting gap. We also find that magnetic
field is suppressed as temperature increases.

\end{abstract}
\end{titlepage}
Single electron tunnelling experiments of Ralph, Black and
Tinkham~\cite{r1} (RBT) initiated interest on the possibility of
superconductivity in a nanometer size Al grain. Although existence
of superconductivity in a very small metallic particle has been
discussed long time ago~\cite{r2} it has been possible to
investigate such small systems only recently with experiments of
RBT. Anderson's criterion, that superconductivity ceases when
discrete energy level spacing becomes comparable with
superconducting gap is correct in general, should be treated
carefully at such small scales.

Instead of macroscopic properties like zero resistance and
Meissner effect, superconductivity in nanometer size metallic
grain manifests itself as an odd-even or number parity effect,
i.e. dependence of physical properties on whether grain has odd or
even number of electrons in it. Evaluation of discrete energy
spectrum using BCS model or exact diagonalization of finite
systems has been the main subject of theoretical
studies~\cite{r21,r22,r23,r24,r25}. (For a review see
Ref.~\cite{r3}).

For parabolic energy dispersion, which is the case for a grain
with perfect symmetry, discrete energy levels are multiply
degenerate~\cite{r4,r5,r6}. Figure 25.1 of Ref.~\cite{r4} and Fig.
1 of Ref.~\cite{r6} show explicitly that for parabolic dispersion
degeneracy of energy levels can be very large. For example, the
number of solutions satisfying
$n^{2}=n_{1}^{2}+n_{2}^{2}+n_{3}^{2}$, all being integer, can
approach to 40 for $n\leq100$. Although disorder and irregular
shape split the degeneracy, as long as they are weak, we can still
talk about near degeneracy. The main point is that typical scale
of dimensional energy quantization
$\delta_{1}=\hbar\upsilon_{F}/L$ is, $L$ being the size of the
grain, much larger than level spacing $\delta_{2} \simeq E_{F}/N$,
where $E_{F}$ is fermi energy and $N$ is the total number of
electrons, even for disordered samples. If attractive
electron-electron interaction is much less than $\delta_{1}$, but
yet larger than $\delta_{2}$, we can approximate the system as a
single, multiply degenerate energy level provided that disorder is
not too strong. A single energy level is certainly drastic
approximation. Our aim is to find out possibility of
superconductivity, or pairing correlations, in this extreme limit.
Physical systems are probably between bulk BCS model and the
nanoscopic (or quantum) limit. By examining the behavior of a
superconductor at both ends we expect to figure out properties of
real systems. Finally, our earlier works~\cite{r4,r5} show that
exact solution of multi-level model agrees very well with single
energy level model in the large level spacing (or nanoscopic)
limit. The formula, obtained in the single level approximation,
for the ratio of the number parity effect parameters for odd and
even cases gives results within one percent neighborhood of the
exact solution when level spacing is about ten times larger than
the superconducting energy gap.

Our starting point is a single, multiply degenerate energy level.
We consider $d$ states and assume that electrons in the same state
are paired. Therefore the model Hamiltonian becomes
\begin{equation}\label{h1}
H=-g\sum_{n,n'=1}^{d}a_{n'\uparrow}^{\dag}a_{n'\downarrow}^{\dag}a_{n
\downarrow}a_{n\uparrow},
\end{equation}
where $a_{n\sigma}^{\dag}$ $(a_{n\sigma})$ creates (annihilates)
an electron in state $n$ with spin $\sigma$. Such a model has
already been used long time ago by Mottelson to study pairing of
nucleons~\cite{r7}. In our case $n$ describes a state where
time-reversed partner is itself. Hence, we assume that electrons
occupying the same state are paired and they can be scattered into
anyone of the unoccupied states together with the same amplitude.
It is an immediate consequence of the form of the interaction that
singly occupied states give vanishing matrix elements. Single
electrons simply block a state from any scattering event.
Therefore, their mere effect is to eliminate certain states in the
Hamiltonian. On the other hand, doubly occupied and empty states
can be denoted in pseudo-spin representation~\cite{r8} which
allows us to diagonalize the Hamiltonian easily.

Pseudo-spins {\bf s} are introduced by the definition
$s_{n+}=a_{n\uparrow}^{\dag}a_{n\downarrow}^{\dag}$ and
$s_{nz}=\frac{1}{2}(a_{n\uparrow}^{\dag}a_{n\uparrow}+
a_{n\downarrow}^{\dag}a_{n\downarrow}-1)$ which allows us to write
the Hamiltonian as $H=-gS_{+}S_{-}$ where ${\bf S}=\sum_{n}{\bf
s}_{n}$ is the total spin and $S_{\pm}$ are corresponding raising
and lowering operators. Finally, when it is written as $H=-g({\bf
S}^{2}-S_{z}^{2}+S_{z})$, we obtain energy eigenvalues as
\begin{equation}\label{e8}
E(s)=-\frac{g}{4}(N-s)(2d-s-N+2)
\end{equation}
where
\begin{eqnarray}
s=\left\{\begin{array}{c}0,2,4,\cdots,2d-N,\cdots,N
\hspace{2cm} \mbox{for N
even}\\ 1,3,5,\cdots,2d-N,\cdots,N \hspace{2cm} \mbox{for N odd.}
\end{array}\right.
\end{eqnarray}
Here, seniority number $s=d-2S$ has been introduced to simplify
the expressions. We note that for a given number of electrons $N$,
$z$-component $S_{z}$ of the pseudo-spin is fixed and hence
eigenvalues are labelled by only one quantum number $S$.
Degeneracy of each $S$, i.e. several different ways of obtaining
the same total spin value, makes the problem nontrivial.
Therefore, we have have to find the total number of states
$\Omega(s)$ having the same seniority number $s$ to evaluate
thermodynamical quantities. Let us start with the ground state
which corresponds to the largest possible total spin $S$ or
equivalently the smallest possible seniority $s$. For even $N$,
$s=0$ state is unique whereas for odd $N$, due to un-paired
electron there is 2$d$-fold degeneracy. Here, we assume that the
grain interacts with a heat bath where spin flip is possible which
gives the factor 2.

Next, we consider the excited states which can be created not only
by pair breaking (single particle excitations) but also by
changing total spin value with fixed number of pairs (collective
excitations). Single particle and collective excitations turn out
to have the same energy spectrum. Hence, the total degeneracy
$\Omega(s)$ can be obtained by summation of degeneracies of each
configuration. Contribution of un-paired electrons to degeneracy
is nothing but the number of combinations of $d$ states taken $s$
at a time. Including spin degree of freedom results in a factor
$2^{s}$. Different ways of obtaining the same total spin $S$ with
$N$ spin-1/2 particles, which is given by
\begin{equation}
\left( \begin{array}{c} N \\ N/2-S \end{array} \right) - \left(
\begin{array}{c} N \\ N/2-S-1 \end{array} \right),
\end{equation}
brings an additional contribution. Total degeneracy is found by
proper use of the two formulae. Summation over all possible
configurations having the same seniority gives us the total
degeneracy
\begin{equation}
\Omega (s)=\sum_{i=0}^{I} \left[ \left( \begin{array}{c} d-s+2i
\\ i \end{array} \right) - \left( \begin{array}{c} d-s+2i
\\ i-1\end{array} \right) \right] \left( \begin{array}{c} d \\ s-2i
\end{array} \right) 2^{s-2i}
\end{equation}
where $I$ is equal to $\frac{s}{2}$, $\frac{(s-1)}{2}$ for even
$N$ and odd $N$, respectively~\cite{r9}.

Before going to evaluate thermodynamical quantities let us
calculate the ground state energy as a function of number of
electrons $N$, for a given degeneracy $d$. Figure. 1 shows $d=$20
case. For $N=$20, the level is half-filled and we reach the
minimum. This is an expected result since both the number of pairs
and the number of the empty states into which they scatter are
maximized to have the optimum energy. Zig-zag structure of the
curve is due to the parity effect. When $N$ is odd there is a
unpaired electron which is neither scattered out and nor allows
any pair to be scattered into the state it occupies. This {\it
blocking effect} effectively decreases the number of states by
one.

Noting the fact that Zeeman energy shifts of discrete levels are
much more important than orbital diamagnetism~\cite{orbit} we can
immediately write down energy eigenvalues in the presence of a
uniform magnetic field as
\begin{equation}
E_{H} = E_{H=0} -\mu_{B}H(n_{\uparrow}-n_{\downarrow}),
\end{equation}
where $n_{\uparrow}$ and $n_{\downarrow}$ denote the number of
spin-up and spin-down electrons, respectively. Although, we find
$E_{H}$ easily, degeneracy expression is no more simple but
instead changes with magnetic field. However, for small number of
states it is not difficult to count the energy levels to find
degeneracy of each energy eigenvalue.

Having degeneracy of each state we can evaluate specific heat
$C_{v}=\frac{\partial \langle E \rangle}{\partial T}$ as a
function of temperature where
\begin{equation}
\langle E \rangle = \frac{\sum_{s} E(s) \Omega(s)
e^{-E(s)/kT}}{\sum_{s} \Omega(s) e^{-E(s)/kT}}.
\end{equation}
Observation of specific heat variation is one of the most direct
ways to find critical temperature $T_{c}$ and magnetic field
$H_{c}$.

Figure. 2 shows variation of $C_{v}(T)$ with temperature for
different number of electrons. Discontinuity in the specific heat, which is
a indication of a second order phase transition in macroscopic
systems, now manifests itself as a relatively fast variation in
$C_{v}(T)$ curve. Absence of a sharp change results in ambiguity
in determination of $T_{c}$. We are going to identify the
temperature at which the maximum of $C_{v}(T)$ curve is reached as
$T_{c}$. We note that the linear background, i.e. contribution to
$C_{v}(T)$ which is of the form $const.T$, is absent. This is
because linear term is a result of Fermi-Dirac distribution and
continuous spectrum near Fermi energy. Since in our case, we have a
single level, $C_{v}(T)$ curve has a flat background.

In Fig. 3 we plot critical temperature for different magnetic
fields as a function of number of electrons. Higher value of
$T_{c}$ for even number of electrons is an indication of number
parity effect which can be defined as
\begin{equation}
\Delta_{P}=\left|E^{2N+1}-\frac{E^{2N}+E^{2N+2}}{2} \right|
\end{equation}
where $E^{2N}$ is the ground state energy for $2N$ electrons.
$\Delta_{P}$ has no $N$ label because for our model it can be
shown that $\Delta_{P}=gd/2$, independent of $N$. However, at
finite temperature $\Delta_{P}$ should be redefined. One
possibility is to replace ground state energies by thermal
averages where we obtain $\Delta_{P}(T)$ as a decreasing function
of temperature~\cite{r9}. We note that for $H=0$, $kT_{c}/g
\simeq2$ while $\Delta_{P}/g=5$ which gives
$2\Delta_{P}/kT_{c}=5$. With increasing $d$ we observe that ratio
decreases. For $d=100$, which is not plotted due to invisibility
small odd-even dependence, $2\Delta_{P}/kT_{c}=4.6$.  Here the
interesting point is that BCS gap equation
\begin{equation}
\Delta_{\textbf{k}}=-\sum_{\textbf{l}}
V_{\textbf{kl}}\frac{\Delta_{\textbf{l}}}{2E_{\textbf{l}}}\tanh
\frac{E_{\textbf{l}}}{2kT}
\end{equation}
can be solved analytically, with $V_{\textbf{kl}}=-g$,
$E_{\textbf{k}}=E$ and $\Delta_{\textbf{k}}=\Delta$, to give
$kT_{c}/g=d/4=\Delta_{0}/2g$, and we find $2\Delta_{0}/kT_{c}=4$
(rather than 3.5).

Finally, in Fig. 4, we plot critical magnetic as a function of
temperature. We attribute the feature at low temperatures to
slightly ambiguity in the definition of critical temperature. In
interpreting $H_{c}(T)$ curve we should keep in the mind that the
critical magnetic field is determined by Zeeman splitting. For a
different geometry, for example for a ring shaped conductor,
orbital coupling can also play an important role.

In conclusion, we can understand basic physics of thermal and
magnetic properties of a nanoscopic superconductor using our very
simple model which can be solved exactly. In the extreme
nanoscopic limit where a single degenerate or nearly degenerate
energy level is left we predict a different gap to critical
temperature ratio. Our results, with increasing number of levels,
approach to exact solution of the BCS gap equation which gives
$2\Delta_{0}/kT_{c}=4$ (rather than 3.5). This suggest the
equivalence between the parity effect parameter $\Delta_{P}$ and
the superconducting gap $\Delta$. Finally, we observe that the
critical magnetic field is suppressed with increasing temperature.
\begin{acknowledgements}
We gratefully acknowledge useful discussions with Prof. I. O.
Kulik.
\end{acknowledgements}

\begin{figure}
\caption{Ground state energy with respect to the number of
particles $N$ in a grain with $d=20$. }
\end{figure}
\begin{figure}
\caption{Specific heat of a grain with $d=100$ and $N=100,50,2$.
Smooth transition occurs from pair-correlated region to normal
region around the maximum of $C_{v}$.}
\end{figure}
\begin{figure}
\caption{Critical temperature with respect to number of particles
for a grain with $d=10$ for different magnetic fields $\mu
H/g=0,2,3$. Number parity dependent $T_{c}$ decreases with
increasing magnetic field.}
\end{figure}
\begin{figure}
\caption{Critical magnetic field with respect to temperature where
$d=10$ and $N=10$. $H_{c}$ is suppressed with temperature.}
\end{figure}

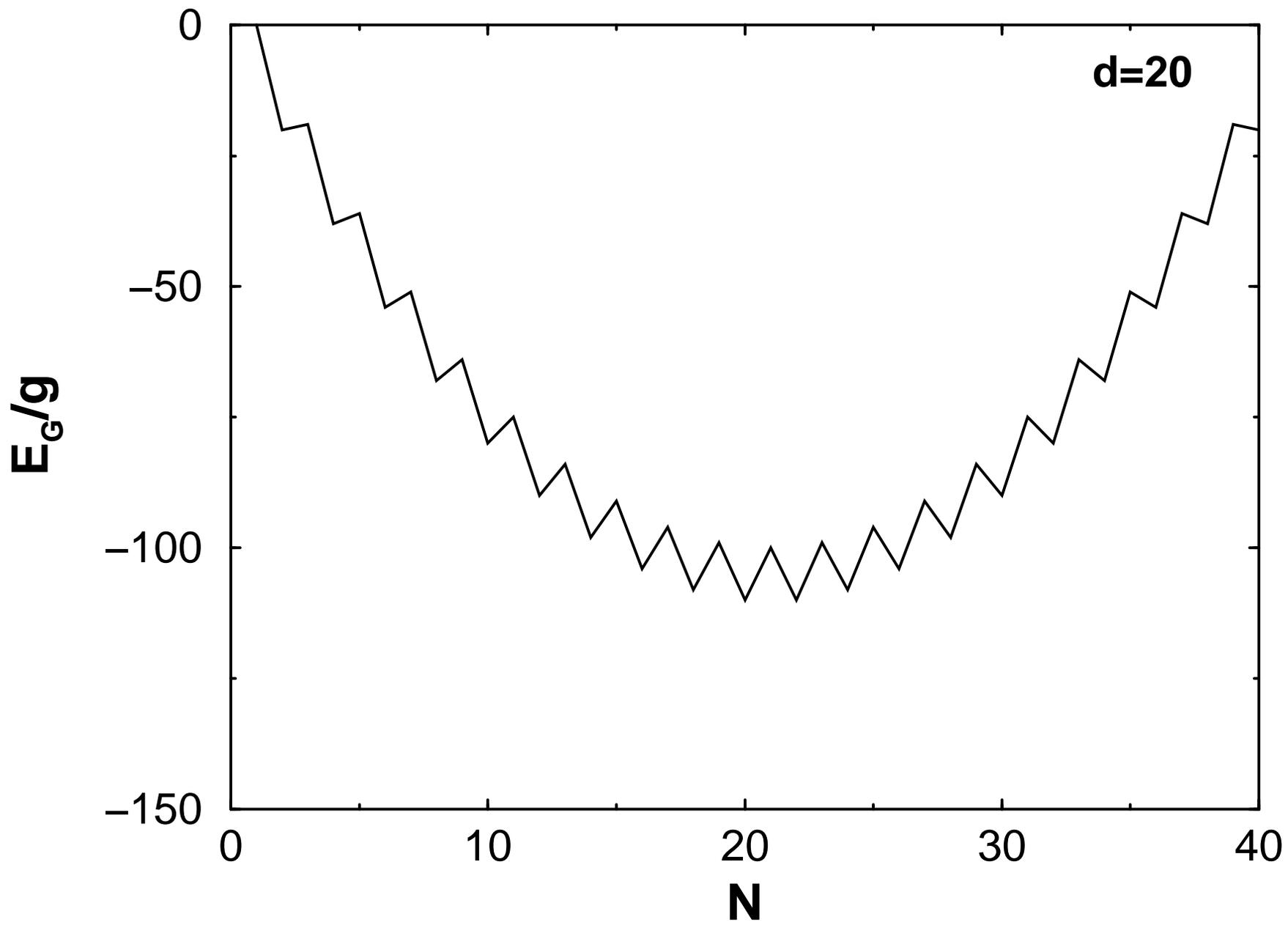
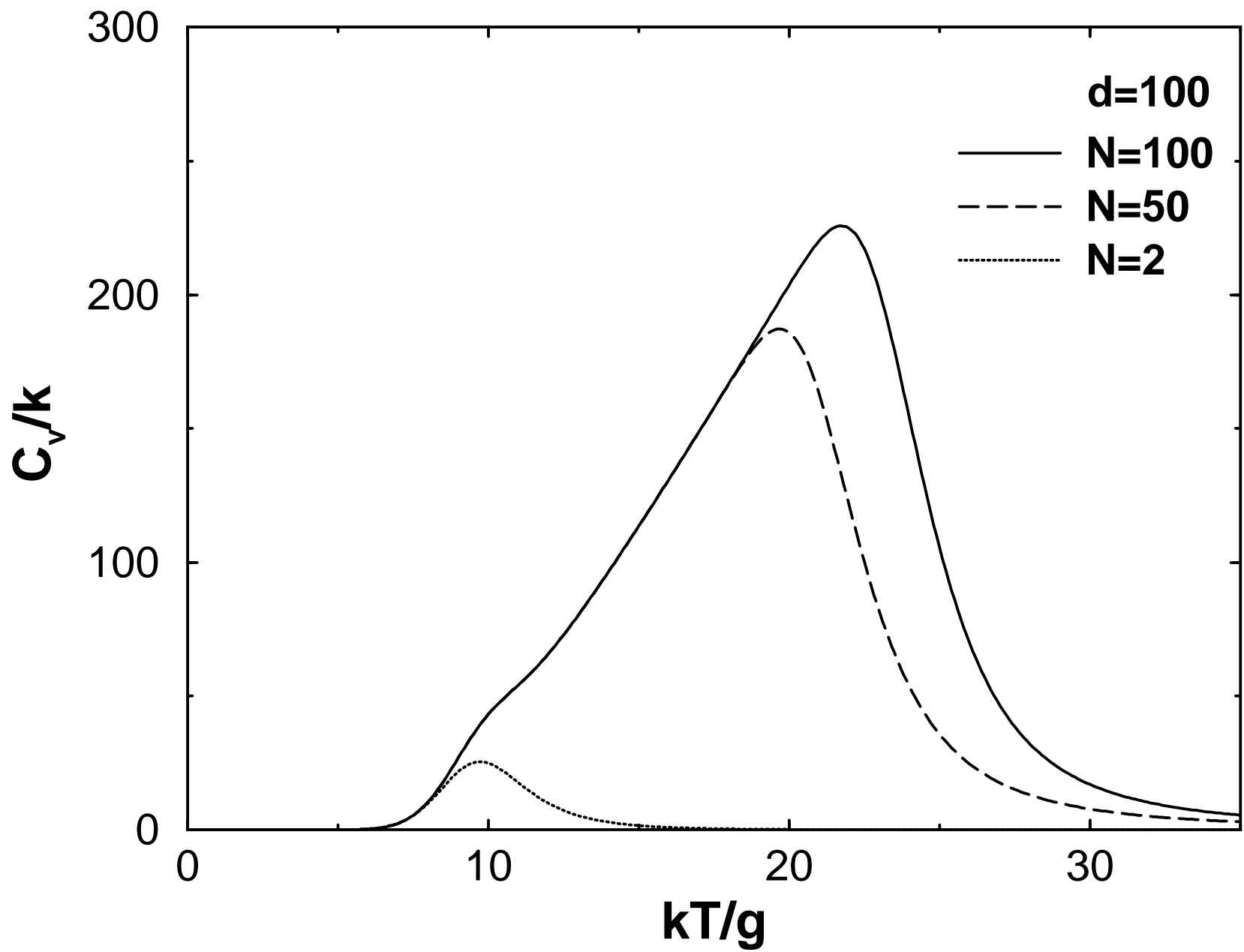
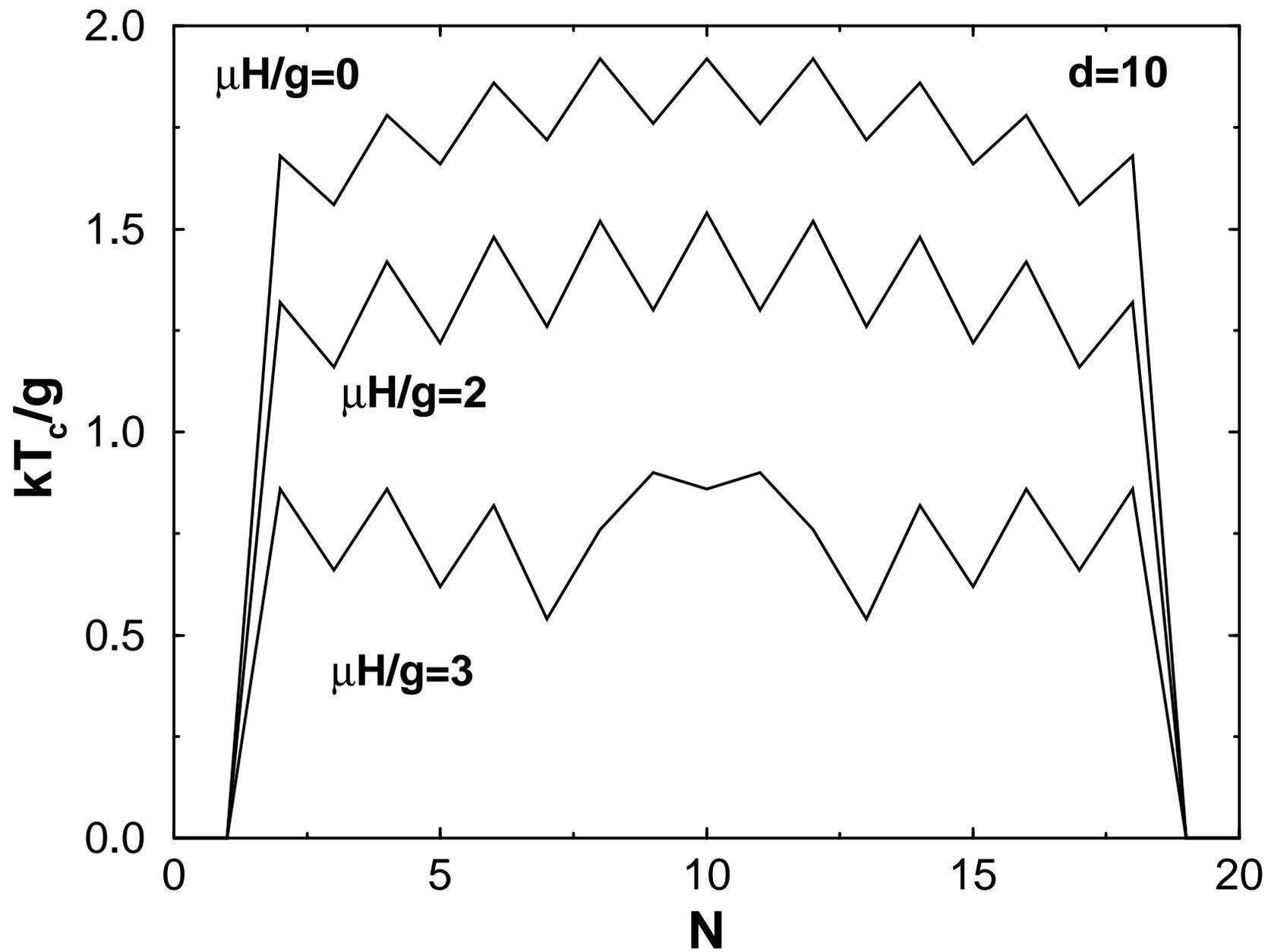
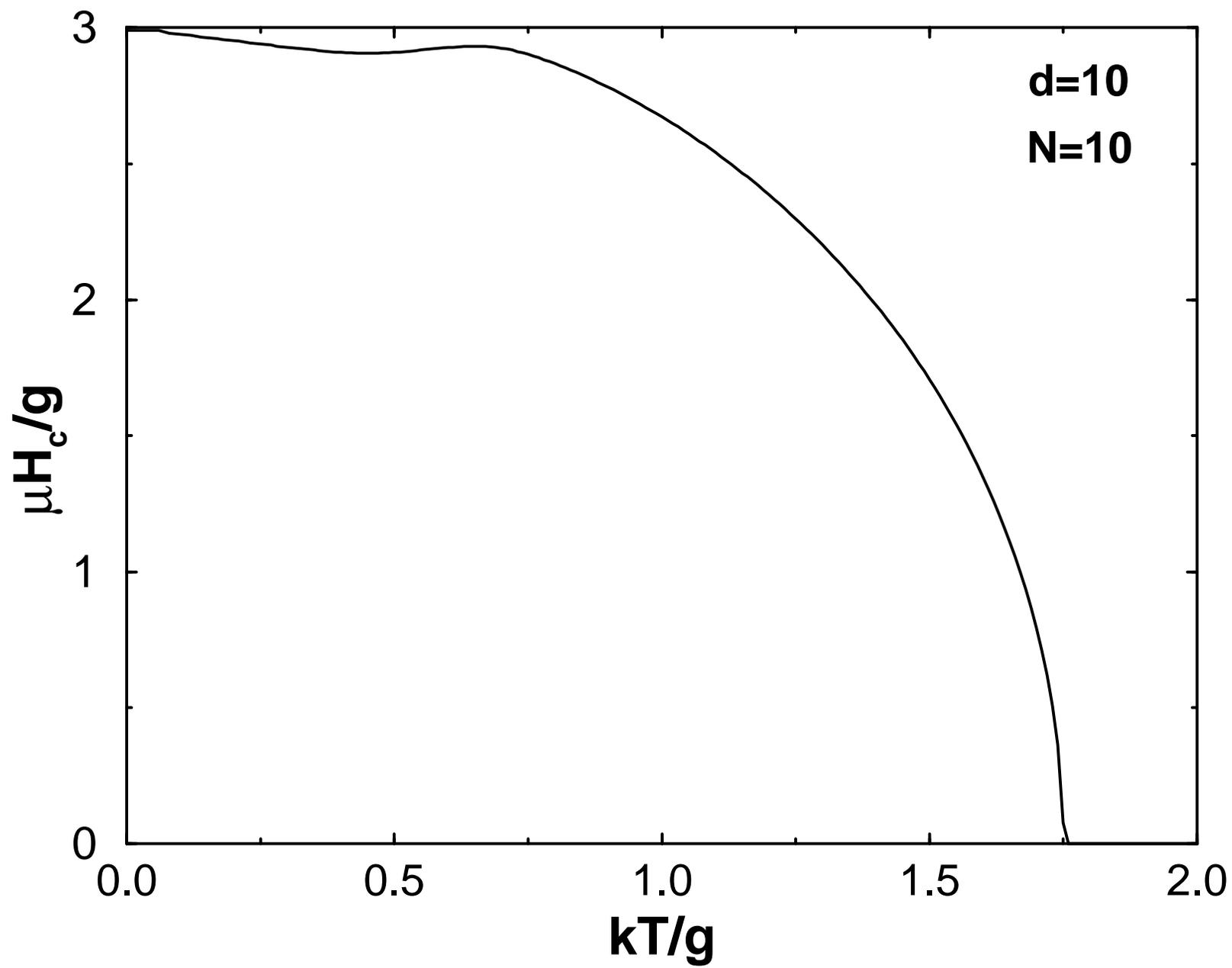
\end{document}